\documentstyle[12pt]{article}
\begin{document}
\newtheorem{thm}{Theorem}
\newtheorem{cor}{Corollary}
\newtheorem{Def}{Definition}
\newtheorem{lem}{Lemma}
\begin{center}
{\large \bf Non-geodesic Motion in General Relativity and
Thermodynamics } \vspace{5mm}

Paul O'Hara
\\
\vspace{5mm}
{\small\it
Dept. of Mathematics\\
Northeastern Illinois University\\
5500 North St. Louis Avenue\\
Chicago, Illinois 60625-4699.\\
\vspace{5mm}
email: pohara@neiu.edu \\}
\end{center}
\vspace{10mm}
\begin{abstract}
In a previous article a relationship was established between the
linearized metrics of General Relativity associated with geodesics
and the Dirac Equation of quantum mechanics. In this paper the
extension of that result to arbitrary curves is investigated. In the
case of scalar fields, a relationship between mass and temperature
is also worked out.
\newline

%\vskip 10pt
\end{abstract}

\section {Introduction}

Before laying out the formalism proper, we need to clarify notation.
Throughout the paper, $({\cal M},g)$ will denote a connected four
dimensional Hausdorff manifold, with metric $g$ of signature -2. At
every point $p$ on the space-time manifold $\cal M$ we erect a local
tetrad $e_0(p), e_1(p), e_2(p), e_3(p)$ such that a point $x$ has
coordinates $x=(x^0,\ x^1,\ x^2,\ x^3)=x^ae_a$ in this tetrad
coordinate system, while the spinor $\psi$ can be written as $\psi
=\psi^ie_i(p)$, where $\psi^i$ represent the coordinates of the
spinor with respect to the tetrad at $p$.  Also at $p$ we can
establish a tangent vector space $T_p(\cal M)$, with basis
$\{\partial_0,
\partial_1,
\partial_2, \partial_3\}$ and a dual 1-form space, denoted by $T^*_p$ with
basis $\{dx_0,dx_1,dx_2,dx_3\}$ at $p$, defined by
\begin{eqnarray}
dx^{\mu}\partial_{\nu}\equiv
\partial_{\nu} x^{\mu}=\delta^{\mu}_{\nu}.
\end{eqnarray}
We refer to the basis $\{dx^0, dx^1, dx^2, dx^3\}$ as ``the basis of
one forms dual to the basis $\{\partial_0, \partial_1, \partial_2,
\partial_3\}$ of vectors at $p$.''

With notation clarified, we note that in a previous paper
\cite{ohara}, the Dirac equation associated with quantum mechanics
was directly related to motion along a geodesic. The linkage was
accomplished in a natural way by associating a generalized Dirac
equation with those operators which are duals of differential
one-forms, obtained by linearizing the metrics of General Relativity
(expressed locally as a Minkowski metric). Specifically if
\begin{eqnarray} ds^2=g_{\mu \nu}dx^{\mu}dx^{\nu}=\eta_{ab}dx^adx^b
\end{eqnarray}
where $a$ and $b$ refer to local tetrad coordinates and $\eta$ to a
rigid Minkowski metric of signature -2, then associated with this
metric and the vector $\bf ds$ is the scalar $ds$ and a matrix
$\tilde{ds}\equiv\gamma_{a}dx^{a}$ respectively, where
$\{\gamma_a,\gamma_b\}=2\eta_{ab}$.

In addition, the $\tilde{ds}$ matrix can be considered as the dual
of the expression $\tilde{\partial}_s\equiv \gamma^a\frac{\partial}
{\partial x^a}$ which in turn enabled us to define a generalized
Dirac equation
\begin{equation} \tilde{\partial}_s \psi\equiv
\gamma^a\frac{\partial \psi} {\partial x^a}=\frac{\partial
\psi}{\partial s},
\end{equation}
associated with the motion of a particle along a geodesic.

This contrasts with the usual method of changing classical
(including special relativity) Hamiltonians into quantum wave
equations. For example, the Dirac equation was first obtained by
substituting the momentum operator for the four-momentum term in the
linearized relativistic Hamiltonian. Similarly, Erwin Schrodinger
used the ``{\it purely} formal procedure''\cite{schr} of replacing
$\frac{\partial W}{\partial t}$ in the Hamilton-Jacobi equation with
$\pm\frac{h}{2\pi i}\frac{\partial }{\partial t}$ to obtain his wave
equation.  In both cases, the transition to quantum mechanics relied
upon additional formal assumptions associated with Hilbert Space
theory, and the final form of the wave equation depended only
indirectly upon the underlying geometry of Minkowski space.

\section{Non-geodesic Motion}

The generalized Dirac equation defined above relies on the
definition of the four-momentum in special relativity and upon the
fact that $\tilde{\partial}_s$ and $\tilde{ds}$ are parallel along
geodesics, and consequently their product is an exact differential
$\frac{d \psi}{ds}ds$. In contrast, when accelerations are
introduced we will find that in general
\begin{eqnarray}\frac{\tilde{ds}}{ds}.\tilde{\partial}_s\psi&=&\frac12\left\{\frac{\tilde{ds}}{ds},\tilde{\partial}_s\psi\right\}+\frac12
\left[\frac{\tilde{ds}}{ds},\tilde{\partial}_s\psi\right]\\
&=&\frac{d \psi}{d s}+\frac{\vec{ds}}{ds}\wedge\vec{\frac{\partial
\psi}{\partial s}},
\end{eqnarray}
and that it is the dot product of $\tilde{ds}$ and its dual
$\tilde{\partial_s} $ that conserve the form of the exact
differential. In addition, no one seems to have noticed that this
term can also be directly related to the Hamilton-Jacobi
characteristic function \cite{synge} associated with a natural
motion, which we now formulate as a Lemma and corollary.

\begin{lem} Let $F(q,t)$ be a function and $\psi(F)
=\exp(kF)$ where $k$ is constant then $p^a=\eta^{ab}\frac{\partial
F}{\partial q^b_i}$ iff $kp^a\psi = \eta^{ab}\frac{\partial
\psi}{\partial q^b_i}$.
\end{lem}
{\bf Proof:} Trivial. It is sufficient to substitute. \vspace{2mm}

\noindent The following corollary immediately follows:

\begin{cor} If $k=1$ and $F=S=\int \eta^{ab} p_adq_b=\int
Hdt-\textbf{pdq}$ is the Hamilton-Jacobi function then
\begin {eqnarray} \gamma^a\frac{\partial \psi}{\partial
x^a}&=&\gamma^a p_a\psi. \end{eqnarray}
\end{cor}
Indeed, the Hamilton-Jacobi function can be directly related to the
metric expressed locally it tetrad coordinates as follows:
\begin{equation} S=\int m\frac{ds}{dt}ds =\int
mdt-\textbf{pdq},\qquad \textrm{ where}\qquad
m=m_0\frac{dt}{d\tau},\end{equation} and chosen units whereby $c=1$
for the velocity of light.

 Equation (6) represents
the most general form of a ``wave-equation'' with respect to a
tetrad coordinate system associated with a particle moving along a
curve with tangent vector $(\frac{dt}{ds}, -\frac{dx^1}{ds},
-\frac{dx^2}{ds}, -\frac{dx^3}{ds})$. In the case of motion along a
geodesic, there exists an eigenvector $\psi$ such that $\gamma^a
p_a\psi=\frac{\partial \psi}{\partial s}=m\psi$ and equation (6)
reduces to the Dirac equation
\begin{equation} \gamma^a\frac{\partial \psi}{\partial
x^a}=m\psi.\end{equation} It is also worth noting that if we take
$k=i=\sqrt{-1}$ that we can also derive the Dirac equation by
considering mass to be a gauge term.

In the case of equation (7), the Hamilton-Jacobi function can be
re-written in covariant form in a general coordinate system as
follows:
\begin{equation} dS=g^{\mu \nu}p_{\mu}dx_{\nu},\end{equation} with
the corresponding wave equation
\begin {eqnarray} \gamma^{\mu}\frac{\partial \psi}{\partial
x^{\mu}}&=&\gamma^{\mu}p_{\mu}\psi \end{eqnarray} associated with
the action along a curve, provided
$2g^{\mu\nu}=\gamma^{\mu}\gamma^{\nu}+\gamma^{\nu}\gamma^{\mu}$.
Also, the Hamilton-Jacobi fuction has the advantage that it
encapsulates information both about the metric $\overrightarrow{ds}$
and its dual $\frac {\overrightarrow {\partial \psi}}{\partial
x^{\mu}}$ in that
\begin{equation} d\psi =\frac{\partial \psi}{\partial x_{\mu}}dx^{\mu}=g^{\mu \nu}\frac{\partial \psi}{\partial
x^{\nu}}dx^{\nu}.
\end{equation}

Moreover from equation (9), we see that the first expression on the
right hand side of equation (4) has ten field terms associated with
the dot product of the tangent vector with the gradient of the wave
function. As a dot product it is the component of
$\tilde{\partial}_s$ along $\tilde{ds}$ which essentially defines
the wave-equation and allows us to clarify the meaning of equation
(4) and generalize equation (8) to
\begin{equation} \gamma^a\frac{\partial \psi}{\partial
x^a}=m\lambda(s)\psi,\end{equation} where
$\lambda(s)=\cos(\theta(s))$ is a directed cosine which varies along
the curve. Note also that when $\lambda(s)=1$ then $\tilde{ds}$ and
its dual spinor $\tilde{\partial}_s$ are strictly parallel and we
get the usual Dirac equation.

Finally, the second term is equivalent to a cross product
$\vec{ds}\wedge\vec{\frac{\partial \psi}{\partial s}}$ of the same
two terms and plays a role similar to $E\wedge B$ in
Electrodynamics. In fact, it is worth noting the similarity between
equation (4) and the famous equation for Lorentz Force on a charge
of size $e$ moving in an electric field $\vec E$:
\begin{equation} \vec {F} = e[\vec{E}+(\vec {v} \times \vec{B})]. \end{equation}

\section {Non-geodesic Motion associated with the Hamiltonian}

In the previous section we related the Hamilton-Jacobi
characteristic function directly to the general form of the wave
equation of quantum mechanics. At the same time because of the
equivalence principle it was noted that the general form of the wave
equation is determined only locally and not globally, especially
when we consider motion along a non-geodesic. Indeed, the existence
of non-geodesics suggests that other factors other than gravity are
at work. The dynamics in such cases is usually analyzed in terms of
test particles. We will continue then for the purpose of this
article to use a somewhat ``classical'' approach to quantum
mechanics, in that we will continue to associate a wave $\psi$ and a
generalized Dirac equation with a particle moving along a curve.
Moreover, from a mathematical perspective $\psi$ can be an $L^p$
function. However, for the purpose of quantum mechanics, we will
take $\psi \in L^2$ or $\psi \in L^2 \otimes H$ where $H$ is a
Hilbert Space.

\subsection {The Physics Interpretation}
Although the wave function can be given a precise mathematical
meaning both as an $L^2$ function and in terms of probability of the
state of the system, from a physics perspective things are more
nuanced. The word ``state'' can be assigned multiple interpretations
depending on the physics of the system and on the question been
asked. Indeed, in Lemma 1 no restrictions were put on the wave
function, other than the fact that it could be written as
$\psi=e^{kF}$. And as it turns out this is a rather weak condition
in that any eigenvector solution to a first order differential
equation must involve the exponential. The state, therefore, may
refer to position, momentum, force, temperature, potential, electric
and magnetic fields etc. In its most general form, we can write
\begin{equation}\gamma^a\frac{\partial
\psi}{\partial x^a}=\phi\end{equation} where $\phi$ would be defined
by the physics of the problem. For example, Maxwell's equations in
Minkowski space can be written in spinor form as
\begin{equation}i\alpha^a\frac{\partial
\psi}{\partial x^a}=-4\pi\phi,\end{equation} where $\phi_0=\rho$ is
charge density, and $\phi_a=j_a,\ a\in\{1,2,3\}$ is a current
density. Also in this case, $\phi_0=0$ and $\phi_a=H_a-iE_a$, where
$H_a$ and $E_a$ are the magnetic and electric fields respectively
\cite{moses}.

For the purpose of this article, we will confine ourselves to
working with eigenvector equations, which means we are essentially
looking for the invariant states of the system. For example, if $A$
is an operator such that $A\psi=\pm \psi$ then $\psi$ can be
interpreted as either an axis of rotation or as the axis of
reflection which remain invariant under the operation. These axes
correspond to the stable states or the states of equilibrium of the
system associated with the operator. One of the challenges then for
physics is to find the equilibrium conditions associated with the
relevant operators (such as the Hamiltonian or Spin operators),
solve their eigenvector equations and then interpret their results.
From a methodological perspective, it should be noted that when
physical states are not in equilibrium or invariant then they are
more difficult to access. This can be seen in the uncertainty
principle, where both the position operator $x$ and the momentum
operators $p_x$ do not have the same eigenfunctions. Consequently,
the physical system cannot be in both invariant states
simultaneously and therefore both cannot be measured at the same
time.

With these observations in mind, we now reconsider Lemma 1 from the
perspective of the Hamiltonian function. Indeed taking our cue from
Hamilton's equations, the canonical equations of motion expressed in
a local tetrad coordinate system are given by
\begin{equation}
\frac{dx^a}{d\tau}=-g^{ab}\frac{\partial K}{\partial p^b}, \qquad
\qquad \frac{dp^a}{d\tau}=g^{ab}\frac{\partial K}{\partial x^b},
\end{equation}
where $K\equiv H\frac{\partial t}{\partial \tau}$ and $H=mc^2$ can
be identified with the Hamiltonian as it appears in the
Hamiltonian-Jacobi function of Equation (7). It now follows from
Lemma 1 that the covariant form of the wave equation associated with
the Hamiltonian and the dual of the metric can be written as
\begin{eqnarray} \gamma^{\mu}\frac{\partial \psi^{\prime}}{\partial
x^{\mu}}&=&\gamma^{\mu}Dp_{\mu}\psi^{\prime}, \end{eqnarray} where
$\psi^{\prime}= \psi^{\prime}(K)$ and
$Dp^{\mu}=\dot{p}^{\mu}+\Gamma^{\mu}_{\nu\eta}p^{\nu}p^{\eta}$. We
also note that both $\psi$ and $\psi^{\prime}$ are not in general
simultaneous eigenvectors of $p^{\mu}$.

For the remainder of this article, we will restrict ourselves to
working with a scalar field in Minkowski space, and in so doing
avoid problems arising from the connection. We will also drop the
prime on $\psi^{\prime}$ and write $\psi$. In the case of a unit
rest mass, Equation (17) then reduces to
\begin{equation} \gamma^a\frac{\partial \psi(s)}{\partial
x^a}=k\gamma_a\left(\frac{d^2x^a}{ds^2}\right )\psi(s).
\end{equation}
Note that this is equivalent to introducing a gravitational
potential of the form
$\ddot{x}^a=\Gamma^{a}_{bc}\frac{dx^b}{ds}\frac{dx^c}{ds}$. Indeed,
the weak field approximation is a special case of equation (18) and
reduces to
\begin{equation} \nabla\psi(s)=k\gamma_a\left(\frac{d^2x^a}{ds^2}\right )\psi(s),
\end{equation}
with the understanding that $\ddot{t}=0$.

In the case of a particle of mass $m(s)$ we can rewrite equation
(19) in the form
\begin{equation}m^2\gamma^a\frac{\partial \psi(s)}{\partial
x^a}=km^2\gamma_a\frac{d^2x^a}{ds^2}\psi(s),
\end{equation}
which in Minkowski space is invariant under Lorentz transformations
and covariant under a change of curve parameter as expressed in the
following lemma:

\begin{lem} Let $\tau$ and $s$ be two parameters of a curve
$\sigma \in ({\cal M},g)$ such that $\frac
{ds}{m(s)}=\frac{d\tau}{M(\tau)}$ along the curve then
$$m^2\gamma^a\frac{\partial \psi(s)}{\partial
x^a}=km^2\gamma_a\frac{d^2x^a}{ds^2}\psi(s)$$ is covariant under a
change of parameter.
\end{lem}
{\bf Proof:} Using $\frac{ds}{m(s)}=\frac {d\tau}{M(\tau)}$, direct
substitution gives:
\begin{eqnarray*} m^2\gamma^a\frac{\partial \psi}{\partial
x^a}&=& km^2\gamma_a\frac{d^2 x^a}{ds^2}\psi\\
&=&kM^2\gamma_a\frac{d^2 x^a}{d\tau^2}\psi
\end{eqnarray*}
and the result follows. \vspace{.2 in}

Finally note that using Special Relativity,we can identify the
parameters $m$ and $M$ with mass. Specifically, if $s$ is the proper
time and $\tau=t$ the local time then
$\frac{dt}{ds}=\gamma=\frac{M}{m}$ where $m$ is the rest mass and
$M=\gamma m$ is the mass in the moving frame.

\subsection {Scalar Fields} It is clear that from a mathematical
perspective there are many possible solutions to equation (18)
depending on the initial conditions. For example, $\ddot{x}^a=-kx^a$
describes simple harmonic motion.  Moreover, under a parameter
change of the form $\frac{ds}{m(s)}=\frac{d\tau}{M(\tau)}$ (see
Lemma),
$p^a=M\frac{dx^a}{d\tau}=M\frac{dx^a}{ds}\frac{ds}{d\tau}=m\frac{dx^a}{ds}$
remains invariant and
\begin{equation} ds^2=\frac{m^2}{M^2}d\tau^2.
\end{equation}

To avoid confusion, let us consider two different parameterizations
for $\sigma$ such that $M=m\frac{ds}{d\tau}$ and $\frac{ds}{d\tau}$
is a variable. Note that it is possible for $M$ to be a variable
along the curve if $m$ is a constant and for $m$ to be a variable if
$M$ is constant along $\sigma$. Equivalently, it is possible to have
two different curves $\sigma_1$ and $\sigma_2$ parameterized by $s$
and $\tau$ respectively such that $m$ is constant along $\sigma_1$
and varies along $\sigma_2$ and vice versa for M. For example, in
the case of particle motion along a curve $\sigma$, parameterized by
a parameter $\tau$, where $\tau$ denotes the proper time along
another curve such that
$\dot{x}^a=\frac{dx^a}{d\tau}=\frac{p^a}{M}$, it follows that
\begin{equation}\int_{\sigma} \dot{x}_adx^a= \int_{\sigma} \dot{x}^a\dot{x}_a
d\tau =\int_{\sigma} \frac{p^a}{M}\frac{p_a}{M}d\tau= \int_{\sigma}
\left(\frac{E^2-{p^2}}{m^2}\right)d\tau=
\int_{\sigma}\frac{m^2}{M^2}d\tau
\end{equation} where $m^2=E^2-p^2$ is a dynamical variable along the curve $\sigma$.
In the case that both curves coincide and $ds=d\tau$ then $M=m$.
Moreover, if we consider one of the curves to be a geodesic, then
$\frac{ds}{m(s)}=\frac{d\tau}{M}$ are exact differentials, since M
is a constant along the geodesic. This suggests a possible
relationship to entropy and temperature, which we will make more
explicit in the next section.

Finally, we note that if we choose $\tau$ to be the world time of
Minkowski space then the approach above is equivalent to the
Stueckelberg approach and equation (21) is identical with equation
(14) described by Horowitz in his paper \emph{On the Definition and
Evolution of States in Relativistic Classical and Quantum
Mechanics}.

\subsection {Relationship of Mass to Temperature}

We now solve (18) for a gas of n independent particles. In other
words, we are taking the simplest of all models and considering the
wave-function associated with each particle to be a scalar field.
Specifically
\begin{eqnarray}
m^2\gamma^a\frac{\partial \psi}{\partial
x^a}=-kM^2\gamma_a\ddot{x}\psi,
\end{eqnarray}
implies that
\begin{eqnarray}
m^2\gamma^a\frac{\partial \tau}{\partial x^a}\frac{\partial
\psi}{\partial \tau}=-kM^2\gamma_a\ddot{x}^a\psi.
\end{eqnarray}
Taking the inner product of both sides with
$\gamma_a\dot{x^a}=\gamma_a\frac{dx^a}{d\tau}$ gives
\begin{eqnarray}
m^2\frac{\partial \psi}{\partial \tau}=-kM^2\dot{x}_a\ddot{x}^a\psi,
\end{eqnarray}
Solving for $\psi$ gives
\begin{eqnarray} \psi = e^{-k\frac{E}{T}}\end{eqnarray}
such that $\frac{E}{T}=\int \frac{M^2}{m^2}\dot{x}_a\ddot{x}^a
d\tau$. In particular, if we consider two different
parameterizations along a geodesic (in terms of world time and
proper time) then both $M$ and $m$ are constant and in this case we
can write
\begin{eqnarray} \psi&=&
c\exp\left({\frac{-kM^2}{2m^2}(-\dot{t}^2+\dot{x}^2_1+\dot{x}^2_2+\dot{x}^2_3)}\right).
\end{eqnarray}
Moreover for a system (gas) of $n$ independent (no interactions
between them) identical particles with unit mass (in the $s$ frame)
the wave function denoted by $\psi(x^a_1,\dots ,x^a_n)$ is given by
\begin{eqnarray} \psi=
c\exp\left[{-k\left(\frac{M}{m}\right)\left(\frac{M}{2m}\right)\Sigma_n(-\dot{t}^2+\dot{x}^2_1+\dot{x}^2_2+\dot{x}^2_3)}\right],
\end{eqnarray} which at any given $t$ coincides with a Bose-Einstein distribution for free
particles with
$\psi_t=c\exp\left({\frac{-\frac{1}{2}(M/m)}{kT}\Sigma_n(\dot{x}^2_1+\dot{x}^2_2+\dot{x}^2_3)}\right)$
where $T=\left(kk_{B}\frac{M}{m}\right)^{-1}$ plays the role of
temperature and $k_{B}$ is Boltzmann's constant. Also note that the
above distribution could also be considered to represent a
Boltzmann's distribution if the gas were composed of $n$
distinguishable particles. The difference between the two cases
would be in the normalizing constants.\newline

\noindent {\bf Remarks:} (1): In the above interpretation we have
separated out the variables in the wave function by writing
$\psi(t,x)=\psi(t)\psi_t(x)$. We can consider $\psi(t,x)$ to be
Lorentz invariant but not $\psi_t(x)$. However, this does not
detract from the theory. Rather it indicates the key role of the
observer when it comes to a local interpretation of the physics
phenomena.

(2) If we let $m$ to be the mass along the curve $\sigma (s)$ and
$M$ be the mass defined with respect to the time parameter along a
local geodesic then we can define the temperature $T=k\frac{m}{M}$
and note that if the particle has positive acceleration then the
temperature is rising, if the temperature is decreasing then the
particle has negative acceleration and if there is no acceleration
then the temperature is constant. It also would mean that as a
massive particle approaches the velocity of light ``c'', its
temperature would become infinite.

(3)  If one associates absolute zero with the absence of all motion
with respect to the ``world time'' in the Stueckelberg frame of
motion then at absolute zero all interactions between matter,
including that of the gravitational field, would have to cease. From
the viewpoint of General Relativity this would mean that there is no
mass at absolute zero, and in this sense the above equation is
consistent.

\section {Conclusion} The article set out to explore the
relationship between particle motion and wave equations within the
framework of General Relativity focusing primarily on non-geodesic
motion. As noted in a previous article these wave equations can be
identified with the wave equations of quantum mechanics if the
proper boundary conditions are imposed. In the process, we
established for scalar fields a relationship between mass and
temperature.

In addition, this approach seems to be comparable to the work of
Stueckelberg and Horowitz \cite{hor} on the evolution of states in
relativistic dynamics. Indeed, if $ds$ is considered to be an
independent variable along a curve then
$ds^2=\frac{m^2}{M^2}d\tau^2$, with $M$ and $m$ having the units of
mass as in equation (21), and with $\frac{m}{M}$ being associated
with an increase of temperature per unit mass. It follows that
$ds=\frac{m}{M}d\tau$ is an exact differential along the curve and
can be associated with entropy.

%%%%%%%%%%%%%%%%%%%%%%%% referenc.tex %%%%%%%%%%%%%%%%%%%%%%%%%%%%%%
% sample references
% "engineering"
%
% Use this file as a template for your own input.
%
%%%%%%%%%%%%%%%%%%%%%%%% Springer-Verlag %%%%%%%%%%%%%%%%%%%%%%%%%%

%
% BibTeX users please use
% \bibliographystyle{}
% \bibliography{}
%
% Non-BibTeX users please use

%%%%%%%%%%%%%%%%%%%%%%%%%%%%%%%%%%%%%%%%%%%%%%%%%%%%%%%%%%%%%%%%%%%%%%

%%%%%%%%%%%%%%%%%%%%%%%%%%%%%%%%%%%%%%%%%%%%%%%%%%%%%%%%%%%%%%%%%%%%%%

%\printindex
\end{document}